\documentclass[prb,aps,twocolumn,epsf,psfig]{revtex4}

\newcommand{\beq}{\begin{equation}}
\newcommand{\eeq}{\end{equation}}
\usepackage{graphics}
\usepackage{epsf}
\usepackage{epsfig}
\usepackage{amsmath}
\usepackage{amsfonts}
\usepackage{amssymb}
\usepackage{graphicx}
\usepackage{epsfig}
\usepackage{bm}

\begin{document}

\title{Signature of an antiferromagnetic metallic ground state in heavily electron doped
Sr$_{2}$FeMoO$_6$}

\author{Somnath Jana,$^{1}$ Carlo Meneghini,$^{2}$ Prabuddha Sanyal,$^{3}$ Soumyajit Sarkar,$^{3}$ Tanusri Saha-Dasgupta,$^{3}$ Olof Karis,$^{4}$ Sugata Ray$^{1,5,\star}$}
\affiliation{$^1$Centre for Advanced Materials, Indian Association for the Cultivation of Science, Jadavpur,~Kolkata 700 032, India}
\affiliation{$^2$Dipartimento di Fisica Universit\'a di Roma Tre Via della vasca navale, 84 I-00146 Roma, Italy and OGG-GILDA c/o ESRF, Grenoble, France}
\affiliation{$^3$Department of Condensed Matter and Materials Science, S.N.Bose National Centre for Basic Sciences. JD Block, Sector III, Salt Lake, Kolkata 700098, India}
\affiliation{$^4$Department of Physics and Astronomy, Uppsala University, SE-75121 Uppsala, Sweden}
\affiliation{$^5$Department of Materials Science, Indian Association for the Cultivation of Science, Jadavpur, Kolkata 700 032, India}
\begin{abstract}
Sr$_{2}$FeMoO$_6$ is a double perovskite compound, known for its high temperature behavior. Combining different magnetic and spectroscopic tools, we
show that this compound can be driven to rare example of  antiferromagnetic metallic state through heavy electron doping. Considering synthesis
of Sr$_{2-x}$La$_x$FeMoO$_6$ (1.0~$\le{x}\le$~1.5) compounds, we find compelling evidences of antiferromagnetic metallic ground state for $x\ge$1.4. 
The local structural study on these compounds reveal unusual atomic scale phase distribution in terms of La, Fe and Sr, Mo-rich regions driven by 
strong La-O covalency: a phenomenon hitherto undisclosed in double perovskites. The general trend of our findings are in agreement with theoretical 
calculations carried out on realistic structures with the above mentioned local chemical fluctuations, which reconfirms the relevance of the kinetic 
energy driven magnetic model.
\end{abstract}
\maketitle
PACS number(s): 75.47.Lx, 75.50.Ee, 72.15.Eb, 78.70.Dm, 79.60.Ht


A general relationship between electrical conductivity and magnetism is maintained in strongly correlated electron systems, where ferromagnetism 
accompanies metallic conductivity and antiferromagnetism is associated with insulating behavior. There are hardly few examples of
antiferromagnetic, metallic (AFM-M) transition-metal oxides, but mostly with layered structures, and only one with a three-dimensional perovskite structure.~\cite{cacro3_prl}
Therefore, the recent theoretical proposition of realizing a metallic, AFM ground state in La doped Sr$_2$FeMoO$_6$ (SFMO) double perovskites,
beyond a critical doping of La,\cite{prabs_2009,Sanyal_2010} sparked curiosity. The theoretical
study in terms of {\it ab-initio} calculation as well as solution of model Hamiltonian
proposed that the stability of this AFM phase arises from the same kinetic-energy-driven mechanism
as originally presented~\cite{DD_PRL} for ferromagnetism (FM) in undoped SFMO. According to th theoretical predictions, the AFM-M phase should be observed in Sr$_{2-x}$La$_x$FeMoO$_6$ with $x$ $>$ 1.5, while there could be  coexistence of FM and AFM phases within a doping range of 1.0 $<$ $x$ $<$ 1.5. Interestingly, this range of electron doping in SFMO has never been explored earlier, although lower doped compounds ($x \le$ 1.0) has indeed been investigated before.~\cite{Sanchez_JMC2003,Alamelu,Coey_jap_SFRO,Navarro_prb2003} It is, therefore, highly
desirable to experimentally probe the large La doping regime, which is unfortunately is complicated by the enhanced
Fe/Mo site-disorder and steric effects with electron doping.~\cite{Navarro_prb2003} Nevertheless, it is a worthwhile attempt to explore
this untested regime to probe the curious proposal of existence of AFM-M phase.
Here, we report structural, magnetic and electronic property studies on Sr$_{2-x}$La$_x$FeMoO$_6$ samples (will be termed as La$_x$ from now onwards) with $x$ upto 1.5 (attempt to go beyond $x$ = 1.5 makes the sample impure). Interestingly, the real chemical structure turned out to be dominated by atomic scale phase fluctuation, a very different scenario than the perfect rock-salt ordering or anti-site like disorder picture commonly assumed for double perovskites, except for the very recent report in one another double perovskite compound, LaSrVMoO$_6$.~\cite{Jana_LSVMO,lsvmo_2nd} Measurements on the series of Sr$_{2-x}$La$_x$FeMoO$_6$ samples clearly indicate that the system is at the verge of adopting a metallic, AFM ground state with higher electron doping ($x~\ge$~1.4). It is also observed that the doped system undergo magnetic frustration in the intermediate doping range of 1.0 $< x <$ 1.4. Consequently, we have also carried out further theoretical calculations, considering the realistic chemical structure, which strongly support the experimental observation. {\bf Both experimental and theoretical study point to the cross-over from ferromagnetic to AFM behavior as largely dominated by the electronic changes, establishing the role of kinetic energy driven mechansim.\cite{DD_PRL}}


The five different compositions
of La$_x$ with $x$~=~1.0, 1.1, 1.25, 1.4 and 1.5 were synthesized in polycrystalline
form by conventional solid state synthesis route. The phase purity of the samples
were checked by x-ray diffraction (XRD) using a Bruker AXS: D8 Advance
x-ray diffractometer.
Magnetic measurements were carried out in
a Quantum Design SQUID magnetometer. X-ray absorption spectroscopy (XAS) was carried out in total electron yield mode
at I1011 and D1011 beam lines of the Swedish synchrotron facility MAX-lab, Lund. The x-ray photoelectron spectroscopic (XPS) measurements
were carried out in an Omicron electron spectrometer, equipped with EA125 analyzer and Mg $K_{\alpha}$
x-ray source. Both XAS and XPS data were collected after {\it in-situ} surface cleaning using a diamond
scraper. Mo $K$-edge x-ray absorption fine structure (XAFS) measurements were performed at the BM08-GILDA beamline at ESRF (Grenoble).
We also carried out theoretical calculations in order to confirm the
experimental observation, both in terms of {\it ab-initio} density functional theory (DFT)
as well as model Hamiltonian approaches, constructed out of DFT calculations.

Phase analysis from XRD refinement indicated space group of P2$_1$/n with small monoclinic
distortion, for all the compositions. This is in agreement with the literature where a transition to monoclinic P2$_1$/n symmetry from
tetragonal I4/mmm symmetry has been reported for a doping level of $x$~$\ge$~0.4.~\cite{Sanchez_JMC2003}

\begin{figure}
\begin{center}
\resizebox{7cm}{!}
{\includegraphics*{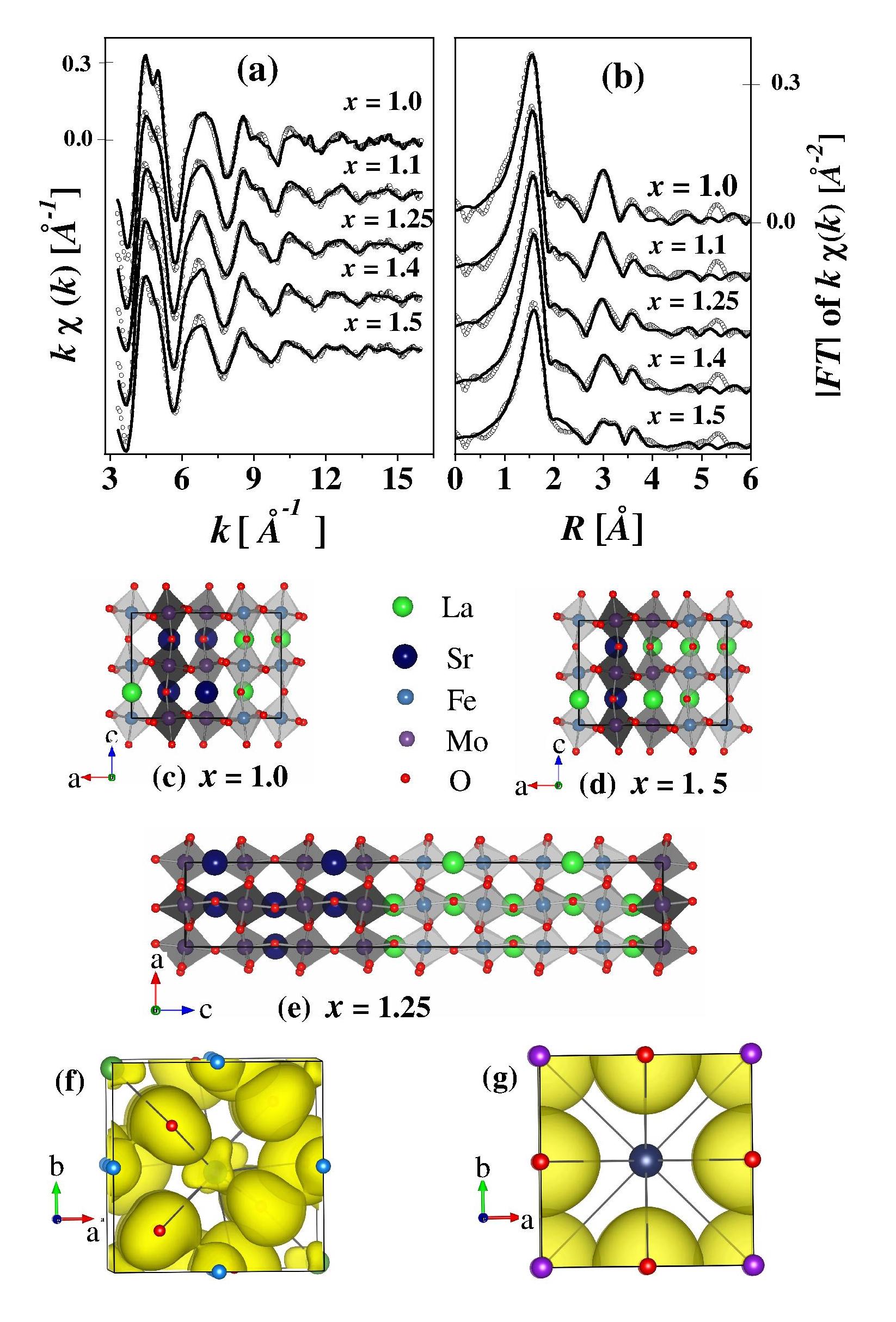}} \\
\caption{(Color Online) (a) The $k$ weighted XAFS data (open circles) plotted together with best fit (black lines).
(b) Fourier transform of experimental (open circles) data plotted together with the fitted (dark lines) curves.
(c)-(e) Representative supercells for $x$= 1.0, 1.5 and 1.25 constructed following the XAFS data. (f) and (g)
Charge density plots for LaFeO$_3$ and SrMoO$_3$, respectively.}
\end{center}
\end{figure}

Firstly, we attempted to probe the true local chemical
structure of the compounds, using XAFS. Figs. 1(a) and 1(b) show the Mo $K$-edge
XAFS data, and the Fourier transform along with the respective best fit spectra for all the compositions (details can be found in the supplementary information (SI)). The most relevant structural information obtained from XAFS analysis concerns the local chemical order around Mo, namely the Mo-O-\emph{B} (\emph{B}=Fe,Mo) and Mo-\emph{A} (\emph{A}=Sr,La) connectivity reported in Table I.
In case of perfectly ordered Fe/Mo arrangement, the number of Mo-O-Mo connections should be 0, for completely random situation, this number should be 3 and for a $A$MoO$_3$-like phase, it should be 6. Interestingly, the XAFS data shows, in all the compounds, the Mo-O-Mo connectivity larger than 3 (random distribution) pointing out a $A$MoO$_3$-rich environment. Similarly, for a random distribution of $x$ La and (2-$x$) Sr atoms in the lattice, each Mo should be surrounded by
(8-4$x$) Sr atoms and 4$x$ La atoms, while experimental data reveals that there is a large preferential accumulation of Sr ions around the Mo sites, signaling the formation of Sr/Mo-rich patches.
\begin{table}[h]
\caption{Mo-$K$ edge XAFS results. N indicates experimentally observed connectivities, while (8-4$x$) is the ideal Mo-Sr connectivity for perfect homogeneous distribution of $A$-site ions. The coordination numbers are constrained to crystallographic structure. The mismatch between experimental data the and best fit is $R^2=0.077$.}
\begin{tabular}{|l|ccc|ccc|}
\hline
& \multicolumn{3}{| c |} {Mo-O-Mo} & \multicolumn{3}{| c |} {Mo-Sr} \\
\cline{2-7}
$x$ & N & R(\AA) & $\sigma^2$ & N (8-4$x$) & R(\AA) & $\sigma^2$ \\
& & & $(\times 10^2$\AA$^2$) & & & $(\times 10^2$\AA$^2$) \\
\hline
1.0 & 4.35 & 3.91 & 0.70 & 6.0 (4.0) & 3.48 & 0.72 \\
1.1 & 4.43 & 3.92 & 0.82 & 6.0 (3.6) & 3.48 & 0.74 \\
1.25 & 5.18 & 3.92 & 1.33 & 5.8 (3.0) & 3.49 & 0.58 \\
1.4 & 4.65 & 3.90 & 0.86 & 5.4 (2.4) & 3.51 & 0.55 \\
1.5 & 4.28 & 3.90 & 0.70 & 4.7 (2.0) & 3.53 & 0.68 \\
\hline
\end{tabular}
\end{table}
Overall, XAFS experiments revealed development of atomic-scale Sr/Mo-rich and consequently, La/Fe-rich patches within the systems, which is very similar to the recent observation in
LaSrVMoO$_6$.~\cite{Jana_LSVMO} The smallest unitcells that satisfy the local Mo-Sr and Mo-O-Mo connectivity for {\it x} = 1.0, 1.5 and 1.25 are shown in Figs. 1(c), 1(d) and 1(e) respectively. It is to be noted that these structures exhibit only representative situation capturing the essential, while the actual structure may be more complex.
{\it Ab-initio} calculation reveals that the formation of patchy structure is driven
by strong La-O covalency, which competes with the stronger Mo-O covalency,
compared to weaker Fe-O covalency,~\cite{lsvmo_2nd} as shown in the calculated charge density plots in Figs. 1(f) and (g).
As a result, it is energetically favorable for La to be in the surrounding of Fe, which
helps to satisfy the covalency between the La and the O, which is connected to two weakly covalent Fe ions.~\cite{lsvmo_2nd} Interestingly, this atomic-scale phase separation further shows a non-monotonic dependence on $x$, with a maxima arising around $x$~=~1.25 (see Table I). This non-monotonic behavior can also be rationalized in terms of stronger La-O covalency.  For a composition with larger $x$, it becomes necessary to accommodate Mo ions at the vicinity of the La ions and within the geometry of growing patchy structure a possibility arises where LaMoO$_3$-like phase develops, which would be highly unfavorable. Therefore, it becomes, preferable for the system to adopt a more homogeneous ionic distribution so that most of the La finds at least some Fe ions around it.

Next, the electronic and magnetic properties of La$_x$ compounds were probed using magnetic,
XAS, and XPS measurements. The magnetization ($M$) {\it vs.} field ($H$) data at 5~K (see Fig. 2(a))
shows that the magnetic moment at higher fields (5 Tesla) is significantly less compared to
what is expected from a perfectly ordered, ferrimagnetic sample. This behavior is consistent with the literature, at least for the La$_{1.0}$
sample,~\cite{Sanchez_JMC2003, Narsinga_JMMM2006, Kahoul_JAP2008} and is generally explained by enhanced contribution from superexchange driven Fe-O-Fe AFM interaction, resulting from increased
anti-site disorder with doping.~\cite{Hemery_prb2006, Poddar_JAP2010} However, the trend of the $M$($H$) curves
indicate a sharp change after $x$~=~1.1 and the $M$($H$) curve already becomes nearly linear for $x$~=~1.25, although a
finite hysteresis persists (see inset). The magnetic coercivity gradually decreases with increasing $x$ and the $M$($H$) from La$_{1.5}$ sample
closely resembles an antiferromagnetic $M$($H$) curve. Out of five samples, two sets of ZFC-FC $M$($T$) data from
the two end compositions measured at a field of 200 Oe are presented in Fig. 2(b).
\begin{figure}
\begin{center}
\resizebox{7cm}{!}
{\includegraphics*{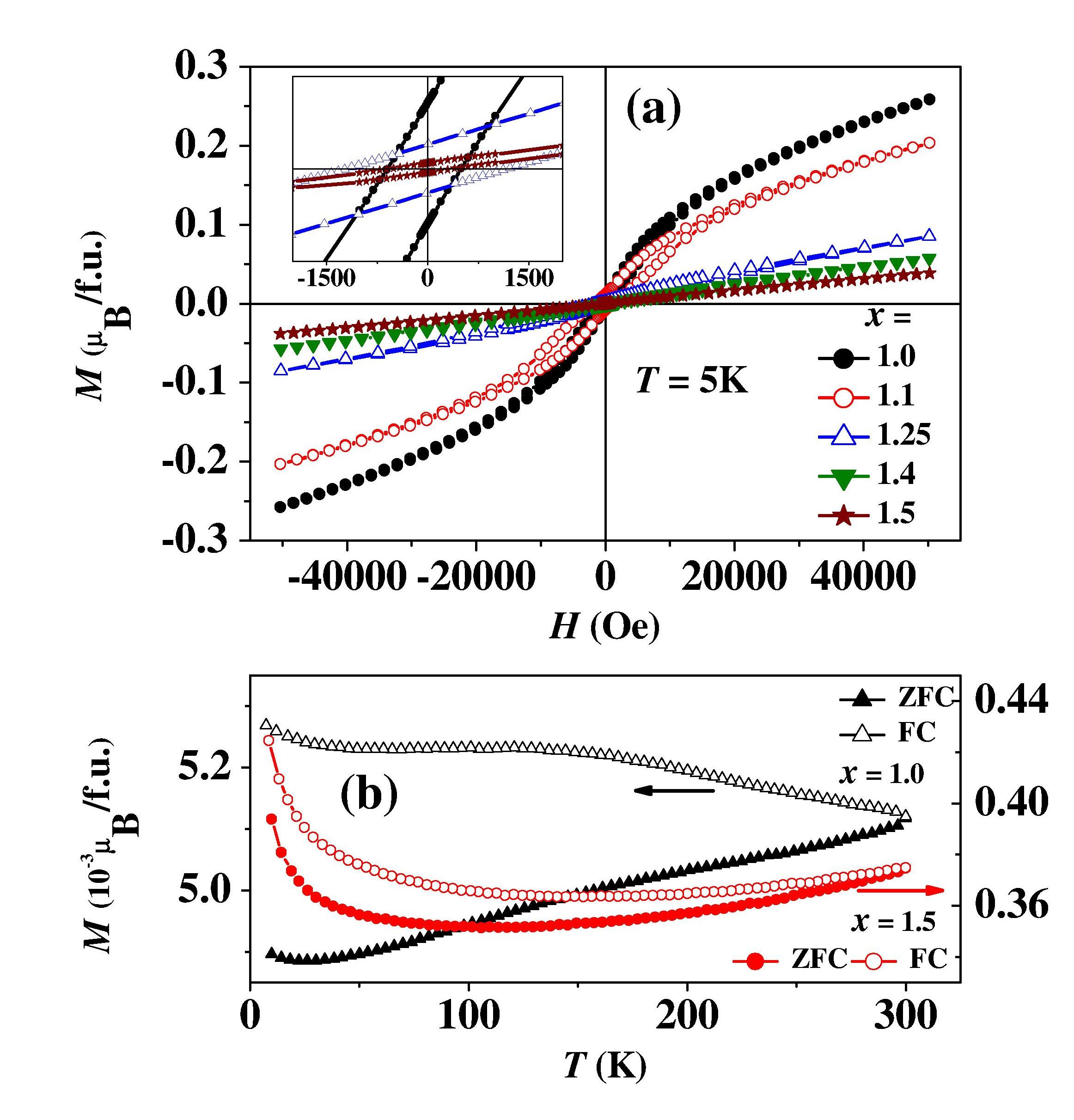}} \\
\caption{(Color Online)(a) Magnetization plotted as a function of varying field for all the compositions. Inset shows the zoomed view of
the plot close to the origin for {~\it x} = 1.0, 1.25, 1.5.
(b) ZFC FC magnetization data plotted as a function of temperature for $x$=1.0 and 1.5 samples.}
\end{center}
\end{figure}
All the ZFC-FC curves exhibit divergences indicating magnetic metastability, presumably originating from the coexistence of FM and AFM interactions. However, there is a gradual reduction in susceptibility as well as in the bifurcation between the ZFC and FC curves, which indicate a steady cross-over towards an antiferromagnetic ground state, though even for the La$_{1.5}$ sample certain magnetic metastability exits.
It is to be noted that the growth of the AFM like magnetic state is not proportional to the available Fe-O-Fe AFM connections, which
in fact decreases at $x>$1.25. {~\it This establishes that the observed crossover in magnetic behavior is entirely driven by changes in electronic structure driven by kinetic energy gain\cite{Sanyal_2010} and is uncorrelated to Fe-O-Fe superexchange, effective in the Fe-rich regions.}

To corroborate the experimental results and also to check the validity of the previous
prediction\cite{Sanyal_2010} of AFM-M phase in La doped SFMO which did not consider the experimentally observed patchy structures,
we carried out further theoretical calculations.
Evidently, several possible supercells with patchy structures could be
constructed with varying distributions of La,~Fe-rich and Sr,~Mo-rich regions, which would be consistent with the XAFS findings.
Unfortunately, it becomes computationally prohibitive to carry out first-principles calculations even for
the simplest supercells (shown in Figs 1(c)-(e)), while in reality, it is expected that the composition fluctuation would be much more random than what
is accommodated in these supercells. To overcome this difficulty, we resort to model Hamiltonian approach,
and introduced the following low-energy, multi-orbital model Hamiltonian, the parameters of which were obtained from {\it ab-initio}
calculations:
\begin{eqnarray}\nonumber
H &=& \epsilon_{Fe}\sum_{i \in \it{B}} f^{\dagger}_{i \sigma \alpha}f_{i \sigma \alpha} + \epsilon_{Mo}\sum_{i \in \it{B'}} m^{\dagger}_{i\sigma \alpha}m_{i \sigma \alpha}\nonumber\\
&&- t_{FM}\sum_{\left\langle ij \right\rangle \sigma, \alpha} f^{\dagger}_{i\sigma \alpha}m_{j \sigma \alpha} - t_{MM}\sum_{\left\langle \left\langle ij \right\rangle \right\rangle \sigma, \alpha} m^{\dagger}_{i \sigma \alpha}m_{j \sigma \alpha}\nonumber\\
&&- t_{FF}\sum_{\left\langle \left\langle ij \right\rangle \right\rangle \sigma, \alpha} f^{\dagger}_{i \sigma \alpha}f_{j \sigma \alpha} + \it{J}\sum_{i \in \it{A}} \mathbf{S}_i \cdot f^{\dagger}_{i \alpha} \vec{\sigma}_{\alpha \beta} f_{i \beta} \nonumber\\
&&+ J_{AS} \sum_{\left\langle \left\langle ij \right\rangle \right\rangle} \mathbf{S}_i \cdot \mathbf{S}_j \nonumber
\end{eqnarray}
\noindent where $f$'s ($m$'s) refer to the Fe (Mo) sites. $t_{FM}$, $t_{MM}$, $t_{FF}$ represent the nearest neighbor
Fe-Mo, Mo-Mo and Fe-Fe hoppings, that happen at the
interface of the patches, and within the Mo-rich and Fe-rich patches, respectively.
$\sigma$ is the spin index and $\alpha$ is the orbital index that spans the $t_{2g}$ manifold of the Fe and Mo $d$
orbitals. The {\bf S$_i$}'s are `classical' core spins at the Fe site, coupled to the itinerant electrons
at Mo site through $J$.~\cite{DD_PRL} The parameter $J_{AS}$ controls the
superexchange driven coupling between Fe spins in Fe-rich patches. The parameters of the model Hamiltonian were extracted~\cite{values}
from the first-principles calculations, through $N$-th order muffin-tin orbital (NMTO) based downfolding calculations,~\cite{tanusri_NMTO} as has been explained in Ref. 2. This model was then solved using exact diagonalization on a patchy supercell, closer to the real structure. The total energies calculated considering the FM as well as
AFM alignment of Fe spins, are plotted in Fig. 3. Evidently, even in presence of a patchy
structure, the AFM solution takes over the FM solution, beyond a critical value of the
number of valence electrons which translates to the critical concentration of La. We
find that the cross-over happens around the concentration of $x$~=~1.0, with small
energy difference between the two solutions (of the order of few meV), thereby having the possibility of phase coexistence
around the crossover point, which is what is recognized experimentally. Also, it is interesting to note that even the current theoretical calculations predict metallic behavior for the AFM state, as stabilization of
the AFM state is hopping driven. In order to experimentally confirm this, we have then carried out XPS valence band experiments on all the samples.

\begin{figure}
\begin{center}
\resizebox{7cm}{!}
{\includegraphics*{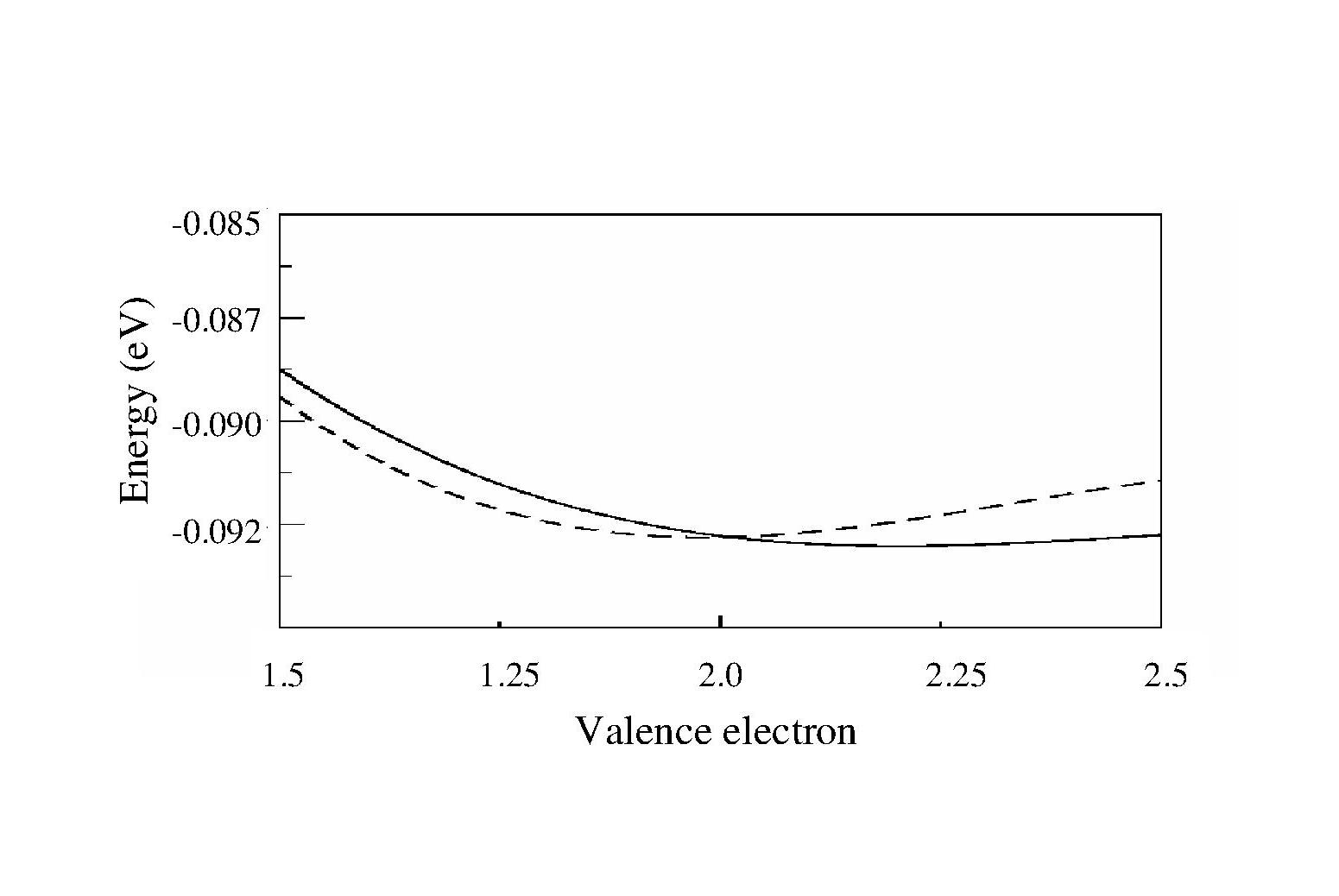}} \\
\caption{Total energies for ferromagnetic (dashed line) and antiferromagnetic alignment of Fe spins, plotted as a function of the number of conduction electrons, as obtained by exact diagonalization of the low-energy Hamiltonian for a 16 $\times$ 4 $\times$ 4 lattice with patchy structure.}
\end{center}
\end{figure}

Valence band spectra, normalized between 5-7 eV (O 2$p$ region) from five samples are summarized in Fig. 4(a). The first important observation is the presence of clear Fermi cutoffs in all the samples, confirming metallicity in all, including the $x$=1.5 sample. This experimental result strongly substantiates the theoretical claim of realizing a AFM-M state with large electron doping in SFMO. It is also important to note that the intensity of the feature just below Fermi energy, constituted by hybridized Fe $t_{2g}$, Mo $t_{2g}$, and O $p$ band for the ferromagnetic case, exhibits a non-monotonic behavior with doping. This intensity increases strongly by going from $x$=1.0 to $x$=1.1, which is expected for electron doping in the ferromagnetic system. However, with AFM behavior gradually taking over as a result of further increase in $x$ (see Fig. 3), the intensity starts to deplete gradually, and a sharp reduction is observed at $x$=1.5, which is consistent with the theoretical prediction of a peak/crest in the FM DOS and a trough in the AFM DOS at large doping.~\cite{Sanyal_2010} The spectrum from $x$=1.5 compound also shows a sudden decrease in intensity at around 8~eV binding energy, resulting a overall narrowing of the valence band. In case of SFMO, this valence band feature has been shown to be a Coulomb correlation driven satellite feature, possessing substantial Fe 3$d$ and Mo 4$d$ contributions,~\cite{prb_us} where the Mo contribution comes mainly through the strong Fe-Mo hybridization as described in the kinetic-energy-driven ferromagnetic mechanism.~\cite{DD_PRL} However, for an AFM ground state, the Mo contribution gets strongly depressed, as has been theoretically shown in case of La$_2$FeMoO$_6$ while going from a FM to AFM-A magnetic structure.~\cite{Sanyal_2010} Therefore, the strong reduction of intensity at around 8~eV binding energy of the valence band only endorses the fact that the AFM interaction indeed dominates in $x$=1.5 compound.
\begin{figure}
\begin{center}
\resizebox{7cm}{!}
{\includegraphics*{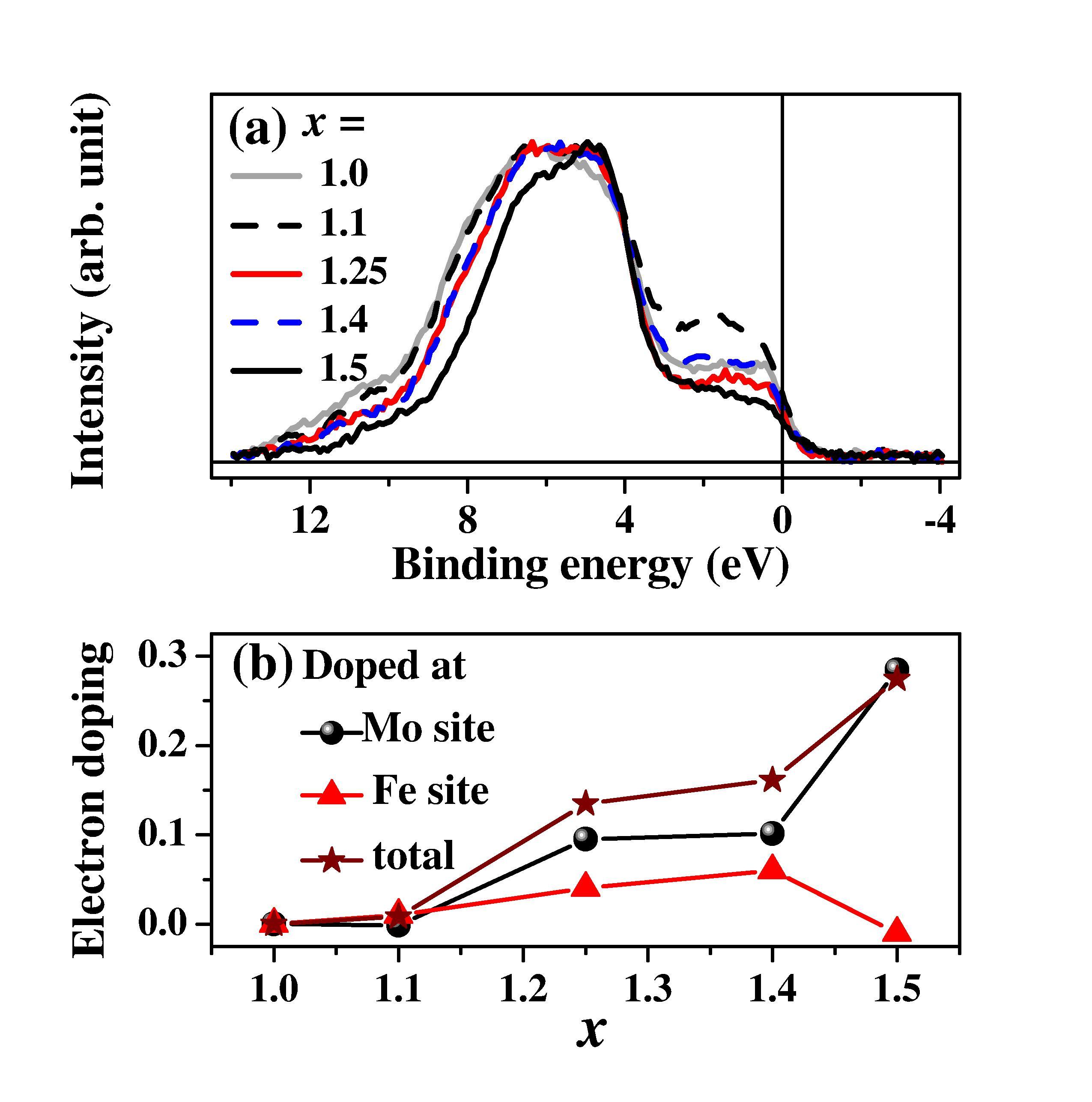}} \\
\caption{(Color Online) (a) Valence band spectra from all the compositions at 300K. (b) The experimental electron doping at the Mo, Fe sites as a function of $x$.}
\end{center}
\end{figure}

We have carried out detailed Fe $L$-edge XAS and Mo 3$d$ core level XPS studies in order to track the location of electron doping. The details of the experiments and the spectra could be found in the SI. In Fig. 4(b) we only show the experimentally obtained variations in Fe, Mo and total charges with doping. It is interesting to note that up to $x$=1.4, the doped electron populates
both Fe and Mo bands but at $x$~=~1.5, almost all the doped electron shifts solely to the Mo site, as was predicted in
the theoretical calculation for ordered double perovskite structure.~\cite{Sanyal_2010}  It can be qualitatively argued that in the
observed patchy structure, La is preferably placed within the cage of Fe-O-Fe. This proximity of La and Fe presumably hinders the transfer
of the doped electrons to the distant Mo site. As the proportion of Fe-O-Mo connectivity increases at $x$~=~1.5, all the doped electrons get transferred to the Mo site.

In summary, the electronic and magnetic structures of La$_{x}$Sr$_{2-x}$FeMoO$_6$ double perovskites
with $x$~$\ge$~1.0 have been studied in detail, for which an unusual AFM-M state was predicted.
XAFS analysis indicates that all the samples contain
small La, Fe rich and Sr, Mo rich patches, originating from strong La-O covalency. Detailed magnetic measurements
provide strong indication of a crossover from a dominant ferromagnetic
to a dominant antiferromagnetic state upon increasing La doping. The XPS valence band shows metallic behavior for all the compounds, while indications of AFM ground state is also revealed at least for $x$=1.5 compound. The experimental results are corroborated with theoretical calculations after taking into account the formation of La, Fe and Sr, Mo-rich short-range patches
within the structure. The theoretical calculations confirm that the stability of AFM-M phase persists even in presence
of the local chemical fluctuation. {\bf Interesting enough, our combined experimental and theoretical study point to the role of kinetic energy
driven mechanism for the enhanced stabilization of AFM in the large doping regime. Our study, therefore, may prompt experimental research along 
similar lines for other classes of double perovskites such as Cr-based 3d-5d compounds,\cite{3d-5d} as well as other systems like pyrochlores\cite{pyro}
and dilute magnetic semiconductors,\cite{dms} for which kinetic energy driven mechanism have been also proposed.}

SJ and SS thank CSIR, India for fellowship. SR thanks DST Fast Track, India for financial support. The work was supported by the Swedish Foundation for International Cooperation in Research and Higher Education. We also thank S. Acharya and D. D. Sarma and their DST SR/S5/NM-47/2005 project for making the photoemission studies possible.


\begin{thebibliography}{99}
\bibitem{values}The values for $t_{FM}$, $t_{MM}$, $t_{FF}$, $\epsilon_{Fe} - \epsilon_{Mo}$ and
$J$ were found to be -0.26 eV, -0.15 eV, -0.40 eV, -0.5 eV, and 1 eV, respectively, while $J_{AS}$ was chosen to be 5 meV, considering the
N\'{e}el temperature of LaFeO$_3$.
\end{thebibliography}
\end{document}




\begin{center}
\bf Supplementary information
\end{center}

\section{Mo $K$ edge X-ray absorbtion fine structure:}
Mo $K$-edge XAFS (x-ray absorption fine structure) measurements were carried out at the BM08-GILDA beamline at ESRF (Grenoble)~\cite{gilda} and the data were analyzed along the lines already described in refs (2, 3), with the aim of understanding, at the local scale, the features of Mo-$A$/$A^\prime$ and Mo-O-$B$/$B^\prime$ disorder as a function of the sample composition. XAFS data, and the Fourier transform along with the respective best fit spectra for all the compositions are shown in Fig. 1(a) and (b) of the paper.

The data refinement required particular attention to take into account the next neighbors distribution as the almost collinear Mo-O-$B$/$B^\prime$ configurations involve non negligible multiple scattering contributions. The standard XAFS formula is used in the fitting. Theoretical amplitude and phase functions were calculated using the FEFF code in muffin-tin approximation using Heidin-Lundqvist inter-atomic potentials.
A representative example of XAFS data fitting is shown in figure 1; the following contributions have been used for all the samples: the MoO, which  represents the contribution coming from the six Oxygens directly bonded to the Mo absorber, the MoLa and MoSr due to the Mo-$A$/$A^\prime$ type and MoOMo and MoOLa coming from the Mo-O-$B$/$B^\prime$ type connections, each one including single (SS) and multiple scattering (MS) contributions.
In order to keep reduced the number of free parameters in data refinement, the multiplicity numbers for the different coordination shells are constrained to the crystallographic values.

The first shell (MoO) contains 6 oxygen neighbors around 2.01 \AA , this distance increases slightly (up to 2.04 \AA ) with raising the La content $x$. More interesting is the evolution of the Mo-$A$/$A^\prime$ and Mo-O-$B$/$B^\prime$ shells. Here the data refinement is achieved fixing the total multiplicity to $N_{\mathrm{Mo}A}=8$ and $N_{\mathrm{MoO}B}=6$ respectively and refining two parameters: $y_A$, being the fraction of Mo-Sr neighbors, and $y_B$ parameter, being the fraction of Mo-O-Mo connections. In this way we obtained the number of Mo-Sr (Mo-La) neighbors: $N_{\mathrm{Mo}A}\times$$y_A$ ($N_{\mathrm{Mo}A}\times$(1-$y_A$)) and the number of Mo-O-Mo (Mo-O-Fe) connections: $N_{\mathrm{MoO}B}\times$$y_B$ ($N_{\mathrm{MoO}B}\times$(1-$y_B$)), which are reported in table I of the paper.

\begin{figure}
\begin{center}
\centering
\resizebox{10cm}{!}
{\includegraphics{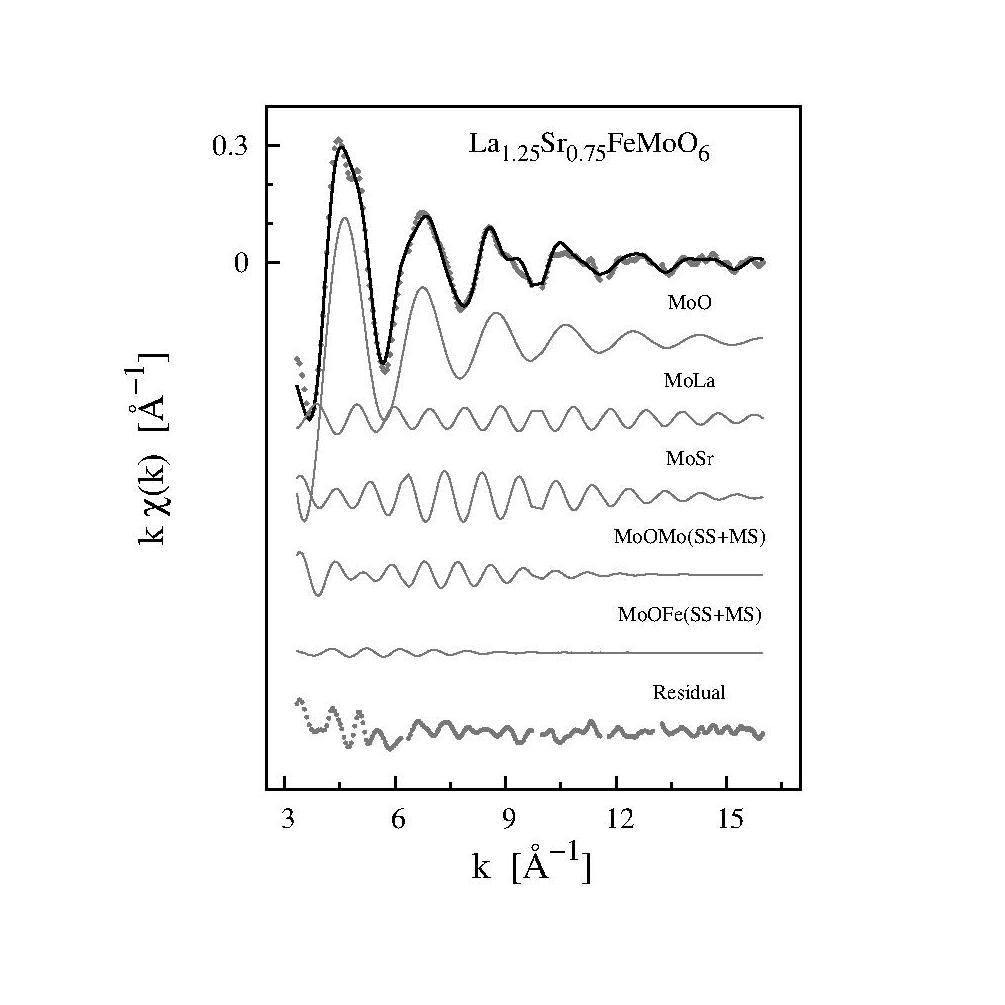}} \\
\vspace{-0.2 in}
\caption{Representative example of XAFS data fitting for the La$_{1.25}$ sample composition: experimental data $k\chi^{exp}$ (points) and best fit $k\chi^{th}$ (full line) are shown at the top and, shifted for clarity, the partial contribution are reported. The lower curve represents the best fit residual: $k\chi^{exp}-k\chi^{th}$.}
\end{center}
\end{figure}

\begin{figure}
\begin{center}
\centering
\resizebox{10cm}{!}
{\includegraphics*{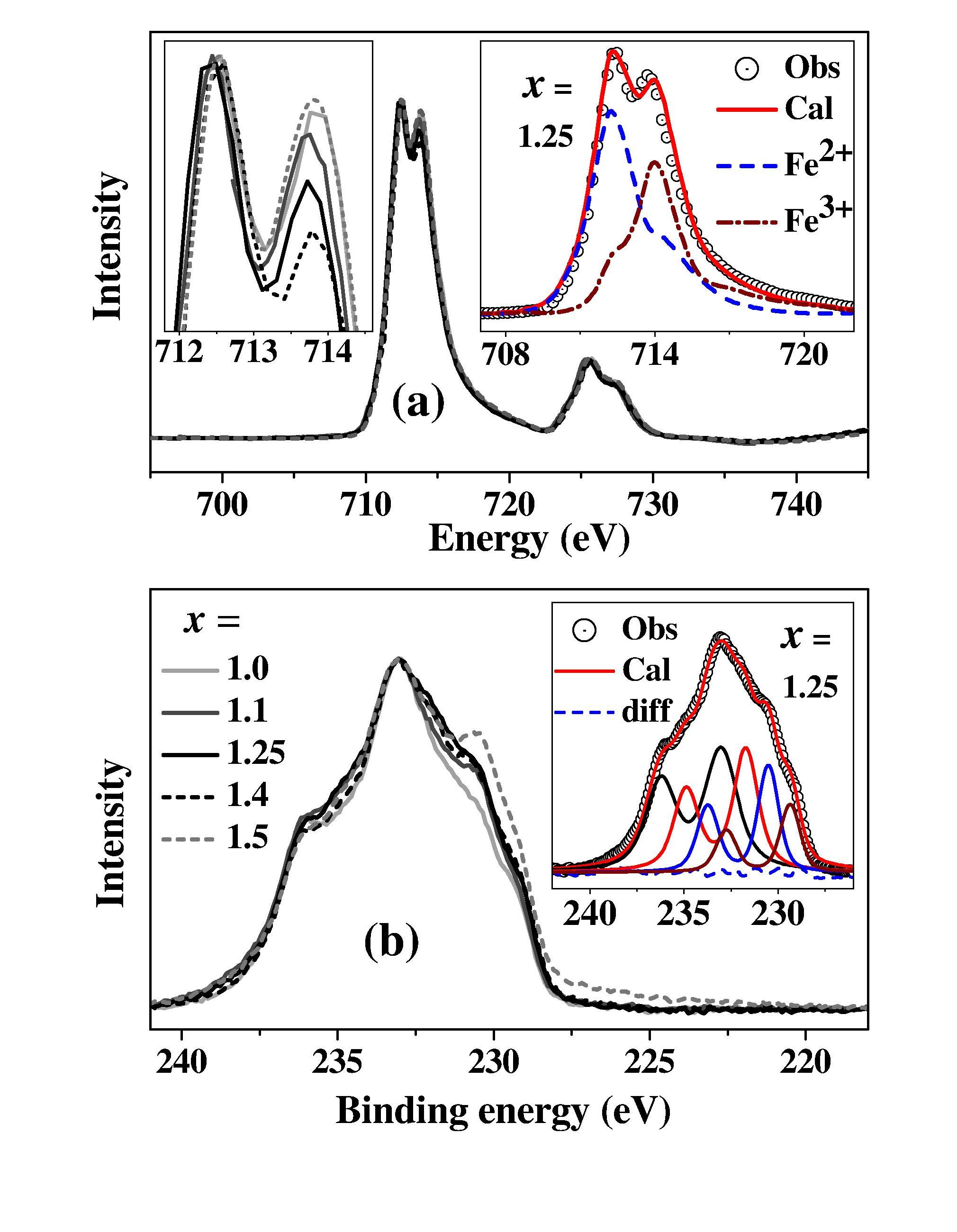}} \\
\vspace{-0.2 in}
\caption{In panel (a) Fe $L$-edge XAS are plotted for all the compositions. Left inset shows the zoomed portion of the 2$p_{1/2}$ and the right inset is the observed and fitted data for $x$=1.25 composition. In panel (b) Mo 3$d$ core level spectra are plotted for all the compositions. Inset shows the representative fitting corresponds to $x$=1.25 composition.}
\end{center}
\end{figure}

\section {X-ray absorbtion spectroscopy (XAS) and x-ray photo electron spectroscopy (XPS):}
According to the theoretical understanding, the doped electrons exclusively go to the Mo site and after a certain doping level above 2.0~e/f.u. ($x>$1.0), the FM state becomes unstable and an AFM state takes over. In order to address the doping trend, we have performed Fe $L$-edge XAS and Mo 3$d$ core level XPS. In the main panel of Fig. 1(a) of the supplementary information all the Fe XAS spectra are plotted. Left inset presents the expanded view of the 2$p_{3/2}$ peak, which closely follows the spectral variation with doping. A regular reduction of the higher energy peak up to $x$~=~1.4, followed by a sudden flip of the trend at $x$~=~1.5 could be observed very clearly. This observation indicates a gradual increase of electron population in the Fe band (Fe$^{3+}$ to Fe$^{2+}$ changeover) upto $x$~=~1.4 and a sharp decrease at $x$~=~1.5. It is to be noted that this observation has been reproduced under proper $in$-$situ$ cleaning and on different batches of samples. We have fitted the experimental Fe XAS spectra as a linear combination of standard Fe$^{2+}$ and Fe$^{3+}$ spectra, acquired from standard literature.~\cite{FeXAS_standard} The right inset of Fig. 1(a) shows a representative set of observed (open circle) and fitted curves (solid line) together with the related spectral weight corresponding to Fe$^{2+}$ (dashed) and Fe$^{3+}$ (dashed-dot) standard spectra for La$_{1.25}$ sample. The change of charge on Fe site in all the samples with respect to La$_{1.0}$, obtained from this fitting, has been shown in Fig. 4(b) in the paper. In the main panel of Fig. 1(b), Mo 3$d$ core level XPS spectra are plotted for the whole composition range. Although a monotonic intensity enhancement at lower binding energy as a function of increased electron doping is very clear from the plot, the enhancement is much more prominent for $x$~=~1.5 sample. We have fitted all the data by a linear combination of few spin-orbit split Mo 3$d$ doublets and a reasonable fitting could be carried out only after considering four such Mo signals, separated by 1.2 eV from each other, and each having a spin-orbit splitting of 3.2 eV. The inset of Fig. 1(b) shows the observed (open circle), fitted (solid) and difference (dash) curves together with the corresponding four doublets for $x$~=~1.25 sample. It is rather puzzling to note that the strongest Mo signal comes at a binding energy corresponding to a Mo$^{6+}$ species, which is inconceivable for these compounds.
However, this has been a consistent observation for any Mo-based double perovskites, including the parent SFMO (nominal constituent is Mo$^{5+}$),~\cite{DD_PRL} or even LaSrVMoO$_6$ (nominal constituent is Mo$^{4+}$).~\cite{lsvmo_2nd} The surface of these molybdates are always prone to oxidation which makes Mo$^{6+}$ (4$d^0$) to be the dominant species at the surface and becomes largely visible in XPS, which is known to be a highly surface sensitive technique.~\cite{Navarro_prb2004} Thus, we have calculated the Mo valence excluding the doublet corresponding to the 6+ state, and rather normalized all the spectra with respect to this extrinsic part of the signal which is expected to remain nearly similar for all the samples under identical experimental conditions. The calculated charge at the Mo site for all the samples relative to La$_{1.0}$ are plotted in Fig. 4(b) in the paper.